# Developing Visual Augmented Q&A System using Scalable Vision Embedding Retrieval & Late Interaction Re-ranker

Rachna Saxena
Fidelity Investments
Bangalore, Karnataka, India
rachna.saxena@fmr.com

Abhijeet Kumar
Fidelity Investments
Bengaluru, Karnataka, India
abhijeet.kumar@fmr.com

Suresh Shanmugam
Fidelity Investments,
Bengaluru, Karnataka, India
suresh.shanmugam@fmr.com

## ABSTRACT

Traditional information extraction systems face challenges with text only language models as it does not consider infographics (visual elements of information) such as tables, charts, images etc. often used to convey complex information to readers. Multimodal LLM (MLLM) face challenges of finding needle in the haystack problem i.e., either longer context length or substantial number of documents as search space.

Late interaction mechanism over visual language models has shown state of the art performance in retrieval-based vision augmented Q&A tasks. There are yet few challenges using it for RAG based multi-modal Q&A. Firstly, many popular and widely adopted vector databases do not support native multi-vector retrieval. Secondly, late interaction requires in-memory computation which inflates space footprint and can hinder enterprise adoption. Lastly, the current state of late interaction mechanism does not leverage the approximate neighbor search indexing methods for large speed ups in retrieval process. This paper explores a pragmatic approach to make vision retrieval process scalable and efficient without compromising on performance quality. We propose multi-step custom implementation utilizing widely adopted hybrid search (metadata & embedding) and state of the art late interaction re-ranker to retrieve best matching pages. Finally, MLLM are prompted as reader to generate answers from contextualized best matching pages. Through experiments, we observe that the proposed design is scalable (significant speed up) and stable (without degrading performance quality), hence can be used as production systems at enterprises.

## CCS CONCEPTS

• Information systems • Information Retrieval • Retrieval models and ranking

## KEYWORDS

Retrieval augmented generation, Vision Embedding, Question Answering system, Visual Augmented Search, Scalability

## 1 Introduction

Business documents often encompass multiple modalities, such as tables, charts, graphs, and diagrams. To fully answer any query, it is essential to extract information from all these complex modalities. While LLMs have simplified text processing, each modality within a document required separate handling, such as OCR, document layout detection, and table extraction etc. [1][2][3], the emergence of multi-modal LLMs (MLLMs) resolves the issue of visual document understanding by mapping queries to a given multi-modal context length [4][5]. MLLMs are specialized in handling modalities have simplified this task by combining the capabilities of vision and text transformer models.

ColPali [6], a Vision Language Model (VLM) built upon PaliGemma model [7], is a special multimodal embedding model that accepts both images and text as input and jointly encode them in numeric representation in same space. It achieves this by dividing image into patches and encoded into a high-dimensional embedding space by the vision transformer. Further, the embeddings are projected to lower dimension and provided to the language transformer to obtain patch embeddings in the language model space. During inference, ColPali adopts ColBert's [8] late interaction mechanism, where a user query is converted to embeddings by the language model and matched against document patches to find the best matches.

While VLMs like ColPali excel at handling images and text together, scaling the solution for inferencing becomes challenging due to the late interaction step. As the number of input documents increases, matching query embedding with all the patch embeddings becomes both computationally and memory intensive. This highlights the need to efficiently reduce the number of documents compared to the query.

This paper explores a feasible solution with vision embedding based approximate nearest neighbor (ANN) retrieval as first pass and state-of-the-art ColPali as re-ranker in second pass [9]. This allows re-purposing the traditionally adopted vector database for practical scalable scenario. In our two-step approach, we initially utilize an existing OpenSearch vector database to index flat vison embeddings and retrieve matches based on cosine similarity with query embedding, ensuring that the retrieval quality remains uncompromised. In the second step, we employ the ColPali late interaction step to re-rank the OpenSearch results. The proposed method is not proprietary and can be utilized by anyone with access to OpenSearch or a similar vector database. The remainder of this paper is organized as follows: Section 2 discusses related work in the field, highlighting ongoing effort by commercial solutions. Section 3 details proposed methodology for scaling multimodal Q&A system using existing databases. In Section 4, we present experiments and results on benchmark and dataset. Finally, Section 5 concludes the study.

## 2 Related Work

In many organizations, most of the available data exists in an unstructured format, often containing various complex modalities. If we talk about PDF documents, before the advent of multimodal models, text and metadata extraction from these modalities involved multiple steps. Text chunking required sophisticated handling. Optical character recognition (OCR) was used to identify text in scanned images, while tables were detected and parsed using layout information. Despite having several high-performing libraries, handling multi-page and nested tables remained a challenge. For graphs and charts data, borders and contours were detected using OpenCV library. Treating each modality separately made the preprocessing pipeline complex, time-consuming, and difficult to maintain.

MLLM based RAG pipelines involved combining text, table and image data to generate embeddings within the same semantic space. For modalities like images and tables, descriptions are generated using vision-language models powered by large language models. These descriptions were processed with text data to get overall context of the document. Often, the image descriptions generated by models fail to comprehensively capture all aspects of complex modalities, leading to information loss.

ColPali offers a state-of-the-art solution to the document preprocessing challenges mentioned above. It produces context-based document embeddings for PDF documents containing charts, graphs, tables, and other modalities. A multipurpose RAG pipeline can be implemented using this model. It not only computes document embeddings but also retrieves them based on query matches. Although ColPali's late interaction retrieval mechanism has demonstrated the best Recall@1 results compared to other vision models, scaling this solution is not feasible. Model generates 1,030 vectors, each of 128 dimensions, for each page. It compares patches of all pages with the query and calculates a single score for each page, indicating how relevant the query is to that page. When the number of pages scales up to hundreds of thousands, comparing embeddings becomes computationally intensive. As ColPali generates 2D embeddings for each page, storing these embeddings in vector databases is not straightforward. Many well-known vector databases do not support multi-dimensional vector storage and retrieval process. Various methods and approaches are explored for minimizing the number of patch embeddings require to compare with query embedding. Vespa ai [10] provides multi-vector HNSW indexing support in proprietary solution. Qdrant [11] implemented two stage retrieval system in which ColPali embeddings are transformed into sparse vectors by mean or max pooling for first stage retrieval. Byadli [12] is a wrapper around ColPali to optimize late interaction by serializing and indexing patch embeddings. This led us to develop an alternative solution involving the serialization and storage of 2D page embeddings at the patch level in a vector database.

## 3 Proposed Methodology

In this experimental setup, we index thousands of documents of PDF format. These documents contain various modalities, including text, images, graphs, and charts. Since handling multidimensional embeddings in-memory is not feasible, we opted for the OpenSearch vector database to store document embeddings and to perform embeddings retrieval. We use the ColPali vision language model (VLM) to obtain embeddings of image documents. This VLM divides each page into 1,030 patches and creates patch embeddings of size 128. We create an OpenSearch index with metadata information, including the document type and name, page number, and patch number. Each patch embedding, along with its metadata, is stored in the vector database. This metadata uniquely identifies each patch of a page for a given document.

Given an input query, its embedding is calculated using the ColPali model. This embedding is two-dimensional; for example, if the number of tokens in a query is 25, then the embedding shape is (25,128), where 128 is the patch embedding size. For each query token, we retrieve the top matching pages from OpenSearch. Along with the top-scoring page, we also shortlist pages with a similarity score of 90% or higher for the next steps. Once we have the best matching page numbers for all query tokens, a set of page numbers is created, and their embeddings are retrieved from the OpenSearch database. Since we store information in the database at the patch level, OpenSearch returns all patches related to a page in response to a search query with a page number filter. Although we can query the database using a patch number filter, using the page number filter is optimal. OpenSearch may not return all the patches of a page in order, so the search output is first sorted in ascending order of patch numbers, and further re-constructed to get the overall page embedding. For a given query, all the best matching pages are retrieved from OpenSearch, completing the retrieval step. For reranking matched output, ColPali is called with the query and page embeddings. VLM evaluates the query embedding against the page embeddings using late interaction and rank pages in order. At the end of this step, we have the reranked output with best matching page on top and so on.

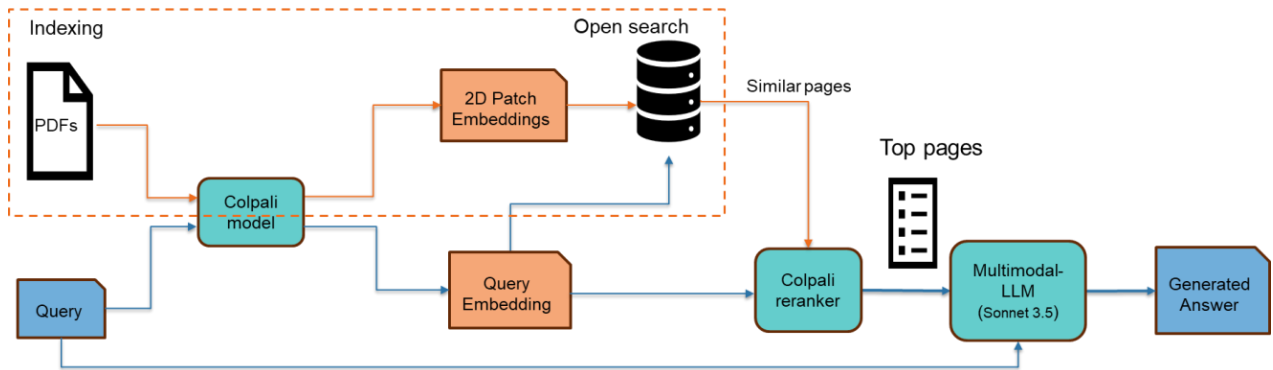

Figure 1: Process flow diagram for Information Extraction from Business Documents.

Once the best matching pages for a query are successfully obtained using the retriever re-ranker pipeline, we utilize prompt template for a LLM (Sonnet 3.5) to extract the exact answer along with cited sources given the matched pages. Section 4. discusses the performance results on datasets for the proposed process in detail.

### 3.1 Developing a scalable indexing mechanism

We utilize an OpenSearch vector database to store flattened patch embeddings obtained from the ColPali model using a HSNW based mechanism. The index also includes storing page number as metadata to support hybrid search on page numbers. For each page, ColPali creates patch embeddings of dimensions (1030, 128), where 1030 denotes the number of patches per page, and 128 is the patch dimension. These patch embeddings are stored in vector DB with metadata to identify each patch uniquely. Following methods depicts a pseudo-code to perform indexing.

```
dict<response> OpenSearchCreateIndex():
    index_name <- unique index name
    #index_body - use hnsw method with cosine similarity algorithm,
    embeddings of type knn_vector with dimension 128
    response= OpenSearch.create(index_name, body=index_body)
    return response

dict<response> OpenSearchIndexData():
    index_name <- unique index name
    file_name <- input file to index
    data_to_index <- get_embeddings_with_metadata(file_name)
    return OpenSearch.bulk(index=index_name, data_to_index)
```

```
list<metadata> get_embeddings_with_meta_data():
    file_name <- input file
    for page in file_name:
        embeddings = ColPali.model(**page) #dim = (1030, 128)
        #create metadata for each patch
        for patch_number from 0 to 1030:
            create metadata dictionary of file_name,
            page number, patch number & embeddings
            append to list_of_metadata
    return list_of_metadata
```

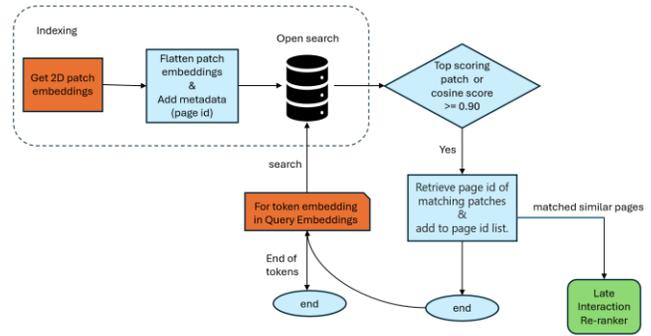

Figure 2: Proposed mechanism using traditional OpenSearch

### 3.2 Retriever and Re-ranker

Figure 2 explains the specific contribution of the literature to implement scalable retriever for vision embedding employing traditional OpenSearch as semantic vector DB. Upon receiving an input query, for each token in query, OpenSearch functions as a retriever, providing the most similar patches along with their corresponding scores as output. Further, page id of matched patches is tracked through a list for further ingestion by re-ranker.

Method retriever: For the given query, embeddings of best matching patches are retrieved from vector DB and rearranged to get full page embedding.

```
list<page_embeddings> retriever ():
    query <- input query
    query_embedding = ColPali.model(**query) #dim = (25, 128)
    for token in query_embedding:
        matching_patches_with_score = OpenSearch.search(token)
        if top_patch_score or patch_score >= 0.9:
            append to list_of_pages
    top_pages = set(list_of_pages)
    for page in top_pages:
        get all the patches corrosponding to page
        append to page_embeddings
    return    page_embeddings
```

Method re-ranker: Once best matching pages are retrieved from vector database, ColPali model is invoked to re-rank them.

```
list<best matching pages> reranker():
    query_embedding <- input query embedding
    page_embeddings <- best matching pages from retriever
    scores = ColPali.evaluate(query_embedding, page_embeddings)
    best_matching_pages = sort(scores)
    return best_matching_pages
```

## 4 Experiments and Results

### 4.1 Dataset

We evaluated the proposed approximate mechanism on two datasets to prove that it performs at par with late interaction mechanism proposed in various works. Firstly, we publish the results on well-known ViDoRe (Vision Document Retrieval) benchmark [13]. Secondly, we evaluated the implemented end to end Q&A pipeline on a proprietary dataset on sustainability report.

1. ViDoRe benchmark: It includes 8 English and 2 French datasets. It encompasses 5 academic and 5 practical tasks [13].
2. ESG Survey dataset: It consists of 150+ questions derived from multi-year sustainability reports of 5 companies. These publicly available reports feature a variety of content, including text, tables, charts, maps, graphs, flowcharts, and diagrams [14][15].

### 4.2 Results and Observations

The results of experiments performed on datasets are presented in this section. Table 1 and Table 2 shows the evaluation metric recall @1 for ViDoRe benchmark and number of pages required for late interaction (indicating low memory usage) respectively.

We evaluated the existing in-memory late interaction for the document retrieval task (ColPali 1.2 version) and compared with the proposed scalable approach i.e. ColPali 1.2 embeddings with OpenSearch retriever & ColPali re-ranker. The results in Table 1. indicate that the proposed approach performed as accurate as existing approach in 8 out of 10 datasets and marginally lower (< 1%) in other two datasets.

Since the proposed approach is an approximation, the best-case performance achieves equal to in-memory ColPali based late interaction retrieval only.

In the existing approach, space complexity is can be approximated as

$$O (q \times p \times m \times d),$$

where q is number of query token, p is the number of pages, m is the number of patches (1030) per page, d is the embedding dimension (128).

As the corpus (p) increases, other variables q, d and m remains range bound (q) or constant (d, m). This is linear complexity with 256 KB memory footprint memory footprint memory per page considering BF16 floating point representation. Since it is must to pass embeddings of all the documents to ColPali model for calculating their relevance with the query, a million pages require 256 GBs GPU memory. When using OpenSearch as the retriever, the first pass retrieval is offloaded to log(n) time complexity with HSNW index-based retrieval [16] without any GPU compute. In second pass, there is significantly reduction in the number of pages passed to in-memory late interaction ColPali re-ranker (few MBs). Table 2. demonstrates the average number of pages used to compare documents with queries.

Table 3. shows accuracy metric (manually validated by sustainability expert) on ESG survey dataset. The pipeline utilized Sonnet 3.5 for final generation task. Experiments with GPT-4o vision model resulted in hallucinated answers specially when answer was not presented. GPT-4o falsely attempted to answer information from document. Sonnet 3.5 v1 correctly response with 'unable to find answer'. Overall, the end-to-end process achieved mean accuracy of 91.8% over 5 clients' survey.

Table 3: Q&A results on ESG dataset

|  | Apple | American Express | Dow | Google | General Motors |
|---|---|---|---|---|---|
| *Pages* | 359 | 123 | 412 | 541 | 89 |
| *Accuracy* | 92.7% | 92.7% | 92.1% | 90.1% | 91.4% |

Table 1: Recall@1 results on ViDoRe dataset

|  | ArxivQ | DocQ | InfoQ | TabF | TATQ | Shift | AI | Energy | Gov. | Health. | Avg. |
|---|---|---|---|---|---|---|---|---|---|---|---|
| *ColPali (+Late Inter.)* | 72.4 | 45.6 | 74.6 | 75.4 | 53.1 | 55.0 | 93.0 | 85.0 | 85.0 | 88.0 | 72.7 |
| *ColPali 1.2 (+Late Inter.)* | 69.2 | 42.4 | 73.2 | 83.57 | 52.61 | 60.6 | 96.0 | 92.0 | 84.0 | 92.0 | 74.56 |
| *ColPali 1.2(+OpenSearch, Late Inter.)* | 68.4 | 42.4 | 73.2 | 83.57 | 52.07 | 62.0 | 96.0 | 92.0 | 84.0 | 92.0 | 74.56 |

Table 2: Number of pages used during late interaction

|  | ArxivQ | DocQ | InfoQ | TabF | TATQ | Shift | AI | Energy | Gov. | Health. |
|---|---|---|---|---|---|---|---|---|---|---|
| *Number of pages in dataset* | 500 | 500 | 500 | 70 | 277 | 1000 | 1000 | 1000 | 1000 | 1000 |
| *ColPali 1.2(+Late Inter.)* | 500 | 500 | 500 | 70 | 277 | 1000 | 1000 | 1000 | 1000 | 1000 |
| *ColPali 1.2(+OpenSearch, Late Inter.)* | 14 | 18 | 44 | 34 | 13 | 53 | 37 | 53 | 41 | 43 |

## 5 Conclusion

This paper proposes a scalable vision embedding retrieval mechanism which can reduce in-memory space requirement by offloading the multi-vector representations to traditional vector databases. It is evident from the experiments on multiple datasets (recall@1) that approximate indexing methods (like HSNW [16]) successfully achieves similar performance utilizing flattened patch embeddings directly and further max-pooled using built-in metadata fields in traditional vector databases. The log(n) order top-k retrieval as first pass can dramatically improve the time efficiency still leveraging the state-of-the-art late interaction mechanism as re-ranker in the second pass. Evaluation of end-to-end visual augmented Q&A process on the ESG survey dataset demonstrate encouraging results. We hope this literature will aid enterprises to implement large-scale visual augmented Q&A system in production with widely adopted traditional vector databases.

**Future Work**. We aim to enhance the time and space efficiency of proposed solution, which we pragmatically demonstrated to be feasible in large-scale settings. Specifically, we intend to extend the experiments by benchmarking pooled patch embeddings in retrieval step and compare performance with current approach to evaluate the magnitude of information loss.

## ACKNOWLEDGMENTS

This work utilized ColPali extensively and extended the same for implementing the proposed solution in a large-scale setting. Our sincere gratitude to the authors and all the contributors of ColPali and related works for the magnificent work.
Disclosure: The opinions provided in this research paper are those of the authors, not necessarily those of Fidelity Investments or its affiliates.